\documentclass[aps,prmaterials,reprint,superscriptaddress,floatfix]{revtex4-2}

\usepackage[T1]{fontenc}
\usepackage[utf8]{inputenc}
\usepackage{lmodern}
\usepackage{textcomp}
\usepackage{graphicx}
\usepackage{amsmath,amssymb}
\usepackage{hyperref}
\usepackage{overpic}
\usepackage{xcolor}
\usepackage{microtype}
\usepackage{placeins}

\newcommand{\affTUD}{New Materials Electronics Group, Department of Electrical Engineering and Information Technology, Technical University of Darmstadt, Merckstrasse 25, 64283 Darmstadt, Germany}
\newcommand{\affBi}{Faculty of Physics, Bielefeld University, Universitätsstrasse 25, 33615 Bielefeld, Germany}
\newcommand{\affWMI}{Walter-Mei{\ss}ner-Institut, Bayerische Akademie der Wissenschaften, Walther-Mei{\ss}ner-Strasse 8, 85748 Garching, Germany}
\newcommand{\DSTU}{Advanced Thin Film Technology Division, Institute of Materials Science, Technical University of Darmstadt, Alarich-Weiss-Straße 2, 64287 Darmstadt, Germany}
\newcommand{\affHZB}{Helmholtz-Zentrum Berlin für Materialien und Energie, 12489 Berlin, Germany}

\begin{document}

\title{Thickness-Dependent Orbital-to-Spin Torque Signatures in Cr/Gd/Co Thin Films}

\author{Tiago de Oliveira Schneider}
\author{Michel Heidkamp}
\affiliation{\affTUD}
\author{Luana Caron}
\affiliation{\affBi}
\affiliation{\affHZB}
\author{Inga Ennen}
\affiliation{\affBi}
\author{Matthias Opel}
\affiliation{\affWMI}
\author{Alexey Arzumanov}
\author{Lambert Alff}
\affiliation{\DSTU}
\author{Markus Meinert}
\thanks{Corresponding author.}
\email{markus.meinert@tu-darmstadt.de}
\affiliation{\affTUD}
\date{\today}

\begin{abstract}
We studied orbital-torque generation in \mbox{Cr(10\,nm)/Gd($t_{\mathrm{Gd}}$)/Co(3\,nm)/TaO$_x$} and the inverted stack \mbox{Co(3\,nm)/Gd($t_{\mathrm{Gd}}$)/Cr(10\,nm)} with \mbox{$t_{\mathrm{Gd}}$} from 0 to $5\,\mathrm{nm}$ by combining electrical harmonic Hall measurements with magnetometry. A detailed understanding of the magnetometric data is obtained by cross-sectional chemical composition mapping. The data show temperature-dependent magnetic compensation points, while elemental analysis provides evidence of pronounced intermixing, in particular of Gd and Co layers. From the harmonic Hall dataset we extract the damping-like (DL) and field-like (FL) torque efficiencies normalized to the applied electric field, $\xi_{\mathrm{DL}}^{E}$ and $\xi_{\mathrm{FL}}^{E}$, and interpret their dependence on the Gd interlayer thickness using two different descriptions: (i) a naive-layer model and (ii) an alloy model that accounts for interfacial mixing. Notably, upon reversing the stack order, the FL contribution changes sign, whereas the DL contribution does not change sign within the harmonic Hall measurements.
\end{abstract}
\maketitle

\section{Introduction}
\label{sec:introduction}

Current-induced spin-orbit torque (SOT) enables the electrical manipulation of magnetization and constitutes a central operating principle of modern spintronic devices. In conventional SOT schemes, a charge current is converted into a transverse spin current, most prominently via the spin Hall effect (SHE), yielding damping-like (DL) and field-like (FL) torque components. The SHE was predicted in early theoretical proposals and was later established experimentally in semiconductors and metals \cite{DyakonovCurrentInducedSpinOrientationOfElectronsInSemiconductors1971,HirschSpinHallEffect1999,katoOberservationOfTheSpinHallEffectInSemiconductors2004,SinovaSpinHallEffects2015}.

More recently, orbital degrees of freedom have been recognized as an additional transport channel: a charge current can generate a transverse flow of orbital angular momentum, even in materials with weak spin-orbit coupling (SOC) \cite{GoOrbitronicsOrbitalCurrentsInSolids,GoIntrinsicSpinAndOrbitalHallEffectFromOrbotalTexture2018}. In transition metals, theory has suggested that SHE and anomalous Hall responses can be rooted in an underlying orbital Hall effect (OHE), where the orbital Hall conductivity can exceed the spin Hall conductivity and remains robust across broad material classes \cite{KontaniGiantOHEInTransitionMetalsOriginOfLargeSpinAndAHE,GoIntrinsicSpinAndOrbitalHallEffectFromOrbotalTexture2018}. This perspective has motivated the emerging field of orbitronics, in which orbital currents act as an efficient source for torques once converted into spin angular momentum \cite{GoOrbitronicsOrbitalCurrentsInSolids}.

Experimentally, orbital-current generation and transport have been demonstrated in a growing set of systems. Magneto-optical detection has provided direct evidence of the OHE in light metals such as Cr or Ti \cite{LyalinMagnetoOpticalDetectionOfTheOHEInCr,ChoiObservationOfTheOHEInALightMetalTi2023}. Complementary electrical approaches have highlighted long-range orbital transport and giant orbital torque efficiencies, with orbital diffusion lengths that can surpass typical spin diffusion lengths \cite{HayashiObservationOfLongRangeOrbitalTransportAndGiantOrbitalTorque2023}. Furthermore, orbital transport parameters such as the orbital Hall conductivity and orbital diffusion length have been quantified, e.g., in V thin films using Hanle magnetoresistance \cite{CasanovaOrbitalHallConductivityAndOrbitalDiffusionLengthOfVanadiumThinFilmsByHanleMagnetoresistance2025}, and giant efficiencies of long-range orbital torque have been reported in bilayers such as Co/Nb \cite{LiuGiantEfficiencyOfLongRangeOrbitalTorqueInCoNbBilayers2023}. These results position 3$d$ transition metals as compelling OHE sources, complementing established heavy-metal SHE platforms.

A key conceptual distinction between spin and orbital transport is that orbital currents do not couple directly to the magnetization. Efficient SOT generation therefore requires orbital-to-spin (LS) conversion, which can take place either in an additional conversion layer or within the magnetic layer itself, provided sufficiently strong SOC is present \cite{DingHarnessingLSConversionOfInterfacialOrbitalCurrentsForEfficientSOT}. Rare-earth materials are particularly promising LS converters because their strong SOC can facilitate an efficient transfer from orbital to spin angular momentum \cite{GambardellaOrbitalTorque,LeeEfficientConversion}. In this context, Cr has emerged as a widely used OHE source, and inserting a rare-earth Gd or Tb layer has been shown to dramatically enhance torque efficiencies \cite{2022SalaGiantOrbitalHall}.

Here, we revisit \mbox{Cr/Gd/Co} trilayers as a model system to investigate orbital-torque generation and detection and its sensitivity to stack order and interfacial chemistry. We study \mbox{Cr(10)/Gd($x$)/Co(3)} and the inverted sequence \mbox{Co(3)/Gd($x$)/Cr(10)} with $x=0\ldots5$~nm, patterned into Hall-bar devices. Throughout this work, thickness values given in parentheses are specified in nanometers. All layer sequences are given in the original deposition order, i.e., substrate / stack / cap. Using electrical second-harmonic Hall (SHH) measurements, we extract DL and FL torque components across the Gd-thickness series, and correlate the torque trends with superconducting quantum interference device (SQUID) magnetometry, ferromagnetic resonance (FMR) measurements and cross-sectional transmission electron microscopy (TEM) with energy-dispersive X-ray spectroscopy (EDX) to assess magnetization reduction and intermixing at the \mbox{Gd/Co} interfaces \cite{andresMagneticBehaviorOfSputteredGdCoMultilayers2002,gonzalezElectricalResistivityAndInterdiffusionInGdCoMultilayers2002,colinoSpinFlopMagnetoresistanceInGdCoMultilayers1999}. By contrasting both stack sequences, we address the sign behavior of DL and FL contributions within the SHH framework and provide an interpretation of orbital-torque generation in light metal/LS conversion layer/ferromagnetic (FM) heterostructures.
To rationalize the experimental trends, we compare two complementary descriptions: (i) a naive-layer model and (ii) an alloy model that accounts for interfacial mixing. 

\section{Methods}

\subsection{Thin-Film Deposition and Lithography}
\label{sec:methods_experiments_deposition_lithography}

The Cr(10)/Gd($t_\mathrm{Gd}$)/Co(3) and Co(3)/Gd($t_\mathrm{Gd}$)/Cr(10) with $t_\mathrm{Gd}$ between 0 and $5\,\mathrm{nm}$ heterostructures were deposited on \mbox{Si} substrates with a \mbox{200-nm} thermal oxide layer using a custom sputter deposition system with 2-inch sources by Bestec GmbH, Berlin. All films were grown at room temperature by DC magnetron sputtering at either \mbox{50\,W} or \mbox{100\,W}. Prior to deposition, the chamber base pressure was \mbox{$5\times10^{-9}$\,mbar}. Ar was used as the sputter gas with a working pressure during deposition of \mbox{$2\times10^{-3}$\,mbar}. The substrates were rotated at \mbox{30\,rpm}, and the angle of incidence was \mbox{$30^\circ$} with a target-to-substrate distance of 12\,cm. After deposition, the stacks were capped with a \mbox{Ta(0.5)/TaO$_x$(1.5)} bilayer. The nominal magnetic layer thicknesses were obtained from the deposition rates determined by a film-thickness monitor, which was previously calibrated with X-ray reflectivity measurements.

\begin{figure}[ht]
  \centering
  \includegraphics[width=\linewidth]{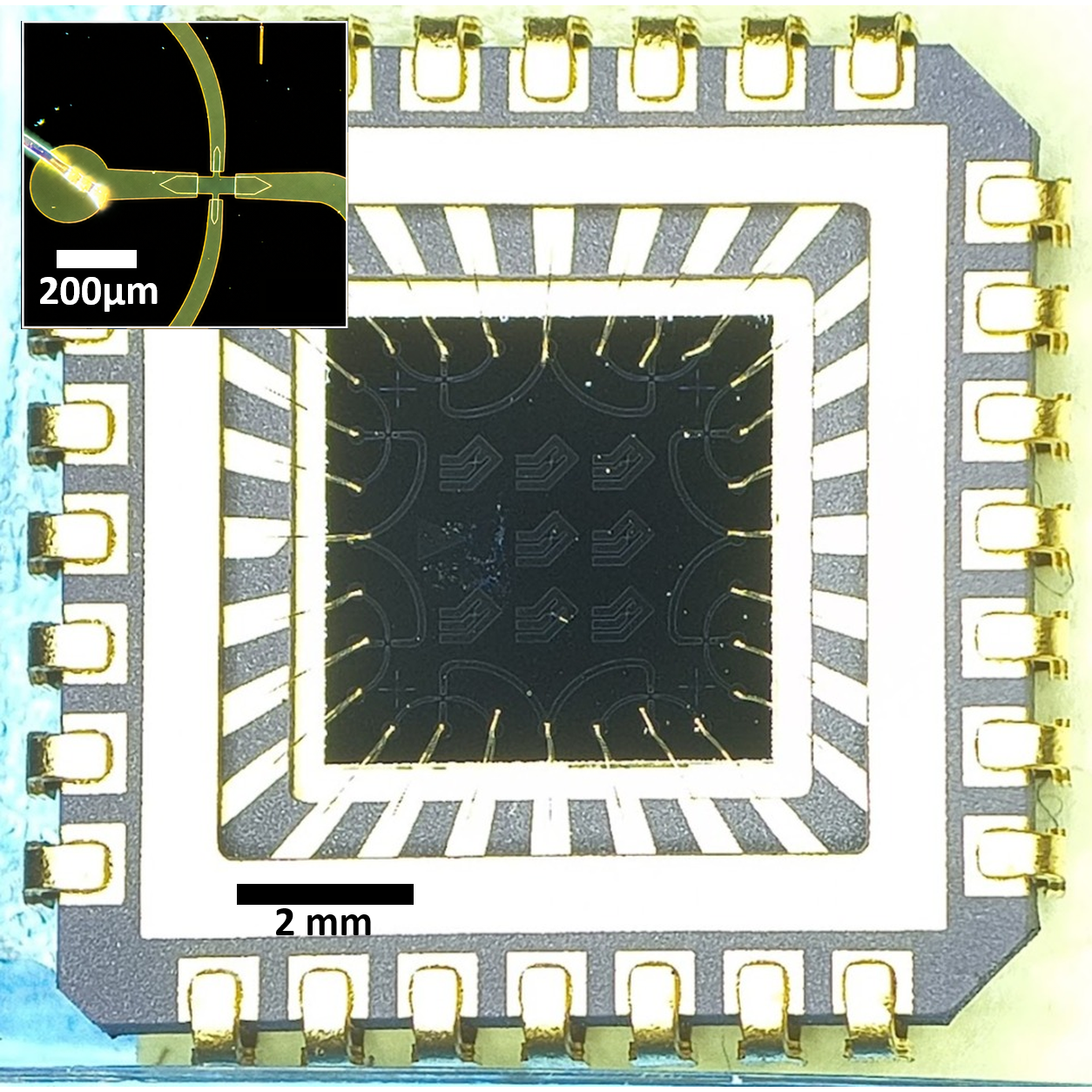}
  \caption{Patterned and wire-bonded sample comprising Hall-bar and ST-FMR devices mounted in a Kyocera CCJ02803 28-pin package. The inset shows a Hall-bar without the complete contact pads.}
  \label{fig:sampleInPackage}
\end{figure}

The Hall-bar devices with a current path width of \mbox{45\,$\mu$m} and length of \mbox{100\,$\mu$m} were fabricated by optical lithography using a Heidelberg Instruments \mbox{$\mu$MLA} maskless aligner. The transverse voltage pickups are \mbox{15\,$\mu$m} wide. Pattern transfer into the film stack was performed by ion-beam etching using a custom \mbox{2-inch} RF ion source operated with an acceleration voltage of \mbox{$V_{\mathrm{acc}}=400$\,V}, a deceleration voltage of \mbox{$V_{\mathrm{decel}}=-150$\,V}, and \mbox{50\,W} RF power.
Electrical contact pads were defined by a lift-off process with an alignment accuracy better than $1\,\mu\mathrm{m}$. To optimize the electrical contacts, a \mbox{15\,s} in situ ion-beam exposure, followed by deposition of \mbox{Cr(10)/Au(50)} without breaking vacuum, was performed. The samples were packaged in 28-pin Kyocera 28 C-QFJ and contacted with $25\,\mathrm{\mu m}$ Au wire as depicted in Fig. \ref{fig:sampleInPackage}. Advantages of this design are the short bond wires and the consistent pin assignment, which facilitates automation of the measurements. Additional spin torque ferromagnetic resonance (ST-FMR) patterns are on the samples but not used for the present study, as we apply the more accurate SHH technique.

\subsection{Magnetic Characterization}

\subsubsection{SQUID Measurements} 

Magnetic characterization was performed using a Quantum Design MPMS3 SQUID magnetometer in VSM mode. The samples were measured with the external magnetic field applied parallel to the plane of the film, i.e. in plane. Temperature-dependent magnetization measurements were carried out in the temperature range from 5\,K up to 300\,K at a fixed in-plane magnetic field of $\mu_0 H = 50\,\mathrm{mT}$. 

For the Cr(10)/Gd(5)/Co(3) sample, the diamagnetic background substration was determined from the high-field region between 5 and $7\,\mathrm{T}$ at $150\,\mathrm{K}$ and kept constant for all other temperatures. For the Co(3)/Gd(5)/Cr(10) sample, the high-field background correction was determined at $5\,\mathrm{K}$. 
For normalization, the samples were diced into $4\,\mathrm{mm} \times 5\,\mathrm{mm}$ pieces using a DISCO wafer saw. The measured magnetic moment was normalized to the corresponding magnetic volume.

\subsubsection{Ferromagnetic Resonance Measurements}
FMR measurements were performed using the OpenFMR broadband spectrometer as described in Ref. \cite{OpenFMR}. The samples were placed on a grounded coplanar waveguide and measured in a field-swept configuration at fixed microwave frequencies between 5 and 27\,GHz. The rectified microwave transmission signal was detected with a Schottky diode and demodulated using lock-in detection with an additional low-frequency field modulation.

For each frequency, the resonance field $B_\mathrm{res}$ and linewidth $\Gamma_B$ were extracted from asymmetric Lorentzian line-shape derivative fits. The frequency dependence of the resonance field was analyzed using the in-plane Kittel relation
\begin{equation}
f_\mathrm{res} = \gamma' \sqrt{B_\mathrm{res}\left(B_\mathrm{res}+\mu_0 M_\mathrm{eff}\right)},
\end{equation}
where $\gamma'=\gamma/2\pi$ is the reduced gyromagnetic ratio and $M_\mathrm{eff}$ is the effective magnetization. The Gilbert damping parameter $\alpha$ was obtained from the frequency dependence of the linewidth according to
\begin{equation}
\Gamma_B(f)=\alpha\frac{f}{\gamma'}+\Delta B(0),
\end{equation}
where $\Delta B(0)$ accounts for inhomogeneous broadening. To improve the fit stability, we fixed the Landé $g$-factor at \mbox{$g = 2.1$}.

\subsubsection{Anomalous Hall Effect Measurements}
The anomalous Hall resistance \(R_\mathrm{AHE}\) and the saturation field \(B_\mathrm{sat}\), required for the SHH analysis, were determined from out-of-plane field sweeps of the first-harmonic Hall voltage, following the procedure commonly used in harmonic Hall analyses \cite{MeinertHallbarGeometry}. To remove voltage offsets, \(R_\mathrm{AHE}\) was extracted from the difference between the saturated Hall voltages at positive and negative magnetic fields,
\begin{equation}
R_\mathrm{AHE}
=
\frac{
V_{1\omega}^{+,\mathrm{sat}}
-
V_{1\omega}^{-,\mathrm{sat}}
}
{2 I_\mathrm{rms}} .
\label{eq:RAHE}
\end{equation}
Here, $I_\mathrm{rms}$ is the root-mean-square value of the applied charge current. The saturation field $B_\mathrm{sat}$ was defined as the field at which $V_{1\omega}$ reaches its saturated value.

\subsubsection{Harmonic Hall Measurements}
The orbital and spin torques were measured with the standard SHH technique \cite{WenTempertureDependenceOfSOTInCuAuAlloys2017,
GarelloSymmetryAndMagnitudeOfSOTInFerromagneticHeterostructures2013}. A Zurich Instruments MFLI lock-in amplifier was used as both the source and the voltage detection unit. The lock-in frequency was set to $2187\,\mathrm{Hz}$ to minimize parasitic noise contributions. The angular rotation of the sample was performed using a stepper motor in increments of \mbox{$10^\circ$}. At each applied magnetic-field value, up to 1.05\,T, a full rotation was measured. To ensure reproducibility and collect measurement statistics, between six and eight Hall-bar devices were automatically measured for each sample. The resulting values are reported as weighted averages and corresponding standard deviation.
The first-harmonic in-phase Hall voltage, arising from the planar Hall effect, is given by
\begin{equation}
V_{1\omega}(\varphi)
=
R_\mathrm{P} I_\mathrm{rms} \sin(2\varphi),
\label{EQU:first_harmonic_hall_voltage}
\end{equation}
where $R_\mathrm{P}$ denotes the planar Hall resistance amplitude, $I_\mathrm{rms}$ is the root-mean-square value of the applied charge current, and $\varphi$ is the angle between the current path and the external magnetic field. The rms current density $j_\mathrm{rms}$ was adjusted in the range of $(3 \dots 5)\times10^{10}\,\mathrm{Am^{-2}}$, depending on the amplitude of the second-harmonic voltage $V_{2\omega}$. For clarity in the following derivation, voltage and angular offsets are neglected in Eqs.~\ref{EQU:first_harmonic_hall_voltage} and \ref{EQU:V2w_decomp}.

\begin{figure}[t]
  \centering
  \includegraphics[width=\linewidth]{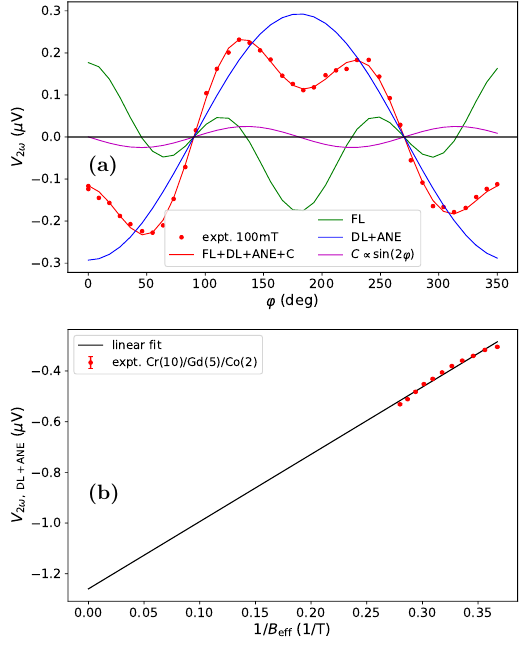}
  \caption{SHH measurement of the \mbox{Cr(10)/Gd(5)/Co(3)} sample. (a) Angular dependence of the SHH signal showing the measured data points (symbols), the total fit (solid red line), and the individual fit contributions assigned to the DL, ANE, and FL terms, as well as an additional residual component $C\propto\sin(2\varphi)$.
(b) DL contribution fit; the error bars are smaller than the symbol size and therefore not resolved.}
  \label{fig:DL}
\end{figure}

The measured second-harmonic Hall voltage $V_{2\omega}(\varphi)$ was fitted using
\begin{equation}
\begin{split}
V_{2\omega}(\varphi)
=
\Bigg[
&
\underbrace{\frac{B_\mathrm{FL}}{B_\mathrm{ext}}R_\mathrm{P}
\cos(2\varphi)\cos(\varphi)
}_{\mathrm{FL}}
\\
&+
\underbrace{
\left(
-\frac{1}{2} R_\mathrm{AHE}
\frac{B_\mathrm{DL}}{B_\mathrm{eff}}
+
\alpha I_\mathrm{0}
\right)
\cos(\varphi)
}_{\mathrm{DL+ANE}}
\\
&+
\underbrace{
C \sin(2\varphi)
}_{\mathrm{residual}}
\Bigg]
I_\mathrm{rms}.
\label{EQU:V2w_decomp}
\end{split}
\end{equation}
Here, $B_\mathrm{ext}$ is the external magnetic field, $B_\mathrm{eff}=B_\mathrm{ext}+B_\mathrm{sat}$, and $B_\mathrm{FL}$ and $B_\mathrm{DL}$ are the FL and DL fields, respectively. The term $\alpha I_\mathrm{0}$ accounts for the anomalous Nernst effect (ANE). The additional $C\sin(2\varphi)$ term captures a residual component, which is reported to originate from a planar Nernst effect driven by an ip temperature gradient \cite{LieAbsenceOfSOTAndDiscoveryOfAnisotropicPNEInCoFeSingleCrystal2023}.

Fig.~\ref{fig:DL} shows a representative SHH measurement of the \mbox{Cr(10)/Gd(5)/Co(3)} sample. In Fig.~\ref{fig:DL}(a), the measured $V_{2\omega}$ data (symbols) are shown together with the total fit (red line) and the individual contributions, with FL in green, DL+ANE in blue, and the residual term in purple. To separate the DL from the ANE contribution, a line fit of the $\cos(\varphi)$ term as a function of $1/B_\mathrm{eff}$ is done: Fig.~\ref{fig:DL}(b) shows this typical fit, where the y-axis intercept corresponds to the ANE contribution and the slope relates to $B_\mathrm{DL}$. 
To avoid uncertainties from the choice of the magnetic layer model and to avoid ambiguities with shunting corrections, we report the measured effective fields normalized to the applied electrical field, $B_\textrm{DL}/E$ and $B_\textrm{FL}/E$. We further report the SOT efficiencies, $\xi_{\mathrm{DL}}^{E}$ and $\xi_{\mathrm{FL}}^{E}$, with respect to the applied electric field $E$ \cite{NguyenSTStudyOfTheSHCAndSpinDiffusionLengthInPtThinFilmsWithVaryingRes2016} within the two magnetic layer models. The efficiencies are defined as
\begin{equation}
\xi_{\mathrm{DL/FL}}^{E}
=
\frac{2e}{\hbar}\,
\frac{M_\mathrm{s}\,t_{\mathrm{FM}}\,B_{\mathrm{DL/FL}}}{E}.
\label{eq:xiDLE}
\end{equation}
Here, $e$ is the elementary charge, $\hbar$ is the reduced
Planck constant, $M_s$ denotes the saturation magnetization of the FM, and
$t_{\mathrm{FM}}$ is the thickness of the FM layer. The extracted values were corrected for the aspect ratio of the Hall-bar geometry \cite{MeinertHallbarGeometry}. In a pure spin Hall system with vanishing FL torque, this quantity approximates the spin Hall conductivity, $\xi_\mathrm{DL}^\mathrm{E} \approx \sigma_\mathrm{SH}$.

\subsubsection{Transmission electron Microscopy}
Samples were prepared for TEM with a focused ion beam system. A JEOL JEM-2200FS was used for high-resolution imaging and elemental composition maps via EDX. 

\FloatBarrier 
\section{Results and Discussion}

\subsection{SQUID Magnetometry}
\label{sec:squid}
\begin{figure*}[t]
  \centering
  \includegraphics[width=\linewidth]{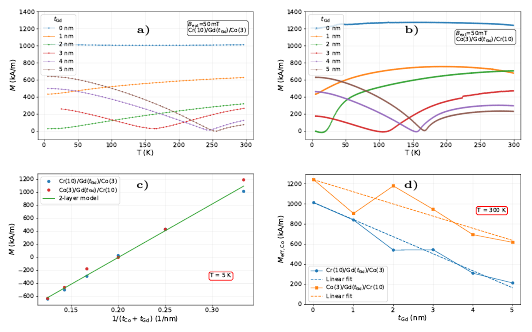}
  \caption{SQUID magnetometry of the (a) \mbox{Cr(10)/Gd($x$)/Co(3)} and (b) \mbox{Co(3)/Gd($x$)/Cr(10)} series with \mbox{$x=0\ldots5$\,nm}. With increasing Gd thickness, the net magnetization is strongly reduced and temperature-dependent magnetic compensation points emerge. The compensation temperature shifts systematically to higher values for larger $x$. The magnetization is normalized to the total magnetic thickness, \mbox{$t_{\mathrm{Co}}+t_{\mathrm{Gd}}$}. (c) Calculated and experimentally determined values of $M_s$ as a function of $1/ (t_\mathrm{Co}+t_\mathrm{Gd})$ at $T = 5\,\mathrm{K}$, assuming a sign reversal at approximately $t_\mathrm{Gd} = 2~\mathrm{nm}$. (d) Magnetization normalized to $t_\mathrm{Co}=3\,\mathrm{nm}$ at $T=300\,\mathrm{K}$. The dashed lines are linear fits to the data.}
  \label{fig:SQUID_MT}
\end{figure*}

Fig.~\ref{fig:SQUID_MT}(a) and (b) summarize the temperature-dependent magnetization $M(T)$ of the \mbox{Cr/Gd/Co} and \mbox{Co/Gd/Cr} series measured at a small in-plane field of \mbox{$\mu_0H=50$~mT} and normalized to the full magnetic layer thickness $t_\mathrm{FM} = t_\mathrm{Gd} + t_\mathrm{Co}$. With increasing Gd thickness, the net magnetization is strongly reduced and distinct magnetic compensation points emerge, where the ferrimagnetically coupled Co and Gd contributions cancel and the net moment approaches zero. The compensation temperature shifts systematically to higher temperatures with increasing $t_{\mathrm{Gd}}$. The observed differences between the stack orders in the low-temperature magnetization are attributed to variations in microstructure and intermixing.

Fig.~\ref{fig:SQUID_MT}(c) illustrates a simple two-layer model in which the net magnetization is approximated as
\begin{equation}
M(T)=
\frac{
M_\mathrm{Co}(T)t_\mathrm{Co}
-
M_\mathrm{Gd}(T)t_\mathrm{Gd}
}{
t_\mathrm{Co}+t_\mathrm{Gd}
}.
\end{equation}
Here, the Co and Gd magnetic moments are assumed to be aligned antiparallel. A negative $M(T)$ means that the Gd moment is dominant. The best linear fit to the magnetization at \mbox{$T=5$~K} is obtained with \mbox{$M_\mathrm{Co}=1100\,kA\,m^{-1}$} and \mbox{$M_\mathrm{Gd}=1650\,kA\,m^{-1}$}. Within this simple model, the net magnetization changes sign at approximately \mbox{$t_\mathrm{Gd}=2$\,nm}, where the Gd contribution starts to dominate over the Co contribution. However, near the compensation point, one expects a spin-flop reorientation of the magnetic moments, which may explain the somewhat reduced magnetizations of the Co and Gd layers, which in fact do not align perfectly antiparallel. The literature values for the magnetization of Co and Gd thin films are $M_\mathrm{s}(\mathrm{Co}) = 1415~\mathrm{kA\,m^{-1}}$ \cite{Ueno2015} and $M_\mathrm{s}(\mathrm{Gd}, 4\,\mathrm{K}) = (2080 \pm 210)\,\mathrm{kA\,m^{-1}}$ \cite{Scheunert2012}. Thus, our findings indicate that both materials do not grow ideally and exhibit probably small grains and substantial disorder. 

At higher temperatures, however, the model no longer captures the measured trends. In particular, since elemental Gd has a Curie temperature of approximately \mbox{$T_\mathrm{C}=293$~K}, such a model would predict a rapid suppression of the Gd contribution near room temperature. In contrast, the measured $M(T)$s show no indication of the Gd phase transition up to \mbox{300~K}. In Fig.~\ref{fig:SQUID_MT}(d) we display the effective magnetization at 300\,K normalized to only the Co thickness $M_\mathrm{eff,Co}$, assuming the Gd contribution is vanishing. Here, we observe a clear decrease of $M_\mathrm{eff,Co}$ with increasing Gd thickness, whereas a naive layer model would result in constant magnetization values independent of the Gd thickness. This discrepancy indicates that the magnetic response cannot be described by two independent, weakly antiferromagnetically coupled magnetic layers. Instead, it points to a strongly coupled and intermixed \mbox{Co--Gd} magnetic region in which the Curie temperature of the Gd layer is substantially enhanced due to Co--Gd exchange interactions. Using the linear fit to $M_{\mathrm{eff,Co}}=(M_{\mathrm{Co}}t_{\mathrm{Co}}-M_{\mathrm{eff,Gd}}t_{\mathrm{Gd}})/t_{\mathrm{Co}}$, we obtain $M_{\mathrm{eff,Gd}}\approx 500\,\mathrm{kA/m}$ for Cr/Gd/Co and $M_{\mathrm{eff,Gd}}\approx 360\,\mathrm{kA/m}$ for Co/Gd/Cr at $300\,\mathrm{K}$.

\begin{figure}[t]
  \centering
  \includegraphics[width=\linewidth]{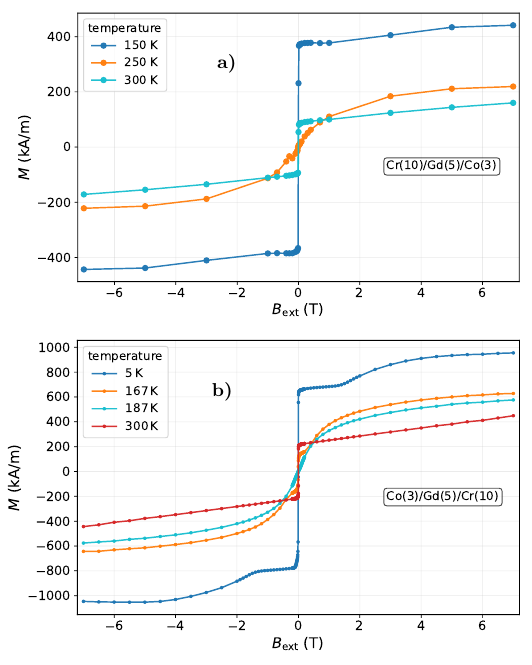}
  \caption{(a) SQUID magnetometry of the $\mbox{Cr(10)/Gd(5)/Co(3)}$ sample and (b) the $\mbox{Co(3)/Gd(5)/Cr(10)}$  sample, showing $B_\mathrm{ext}$ loops measured at different temperatures. The saturation magnetization decreases with increasing temperature. Near the magnetic compensation temperatures of approximately $250\,\mathrm{K}$ and $170\,\mathrm{K}$ for (a) and (b), respectively, the remanent magnetization vanishes.}
  \label{fig:SQUIDCoGdCrMH}
\end{figure}

Representative field-dependent magnetization loops $M(\mu_0H)$ for \mbox{Cr(10)/Gd(5)/Co(3)} and \mbox{Co(3)/Gd(5)/Cr(10)} at selected temperatures are shown in Fig.~\ref{fig:SQUIDCoGdCrMH}. 
Near the compensation points ($\approx250\,\mathrm{K}$  and $\approx170\,\mathrm{K}$ for (a) and (b), respectively), the remanent magnetization is suppressed and the response becomes nearly linear around zero field, consistent with vanishing net moment and a spin-flop reorientation. At lower temperatures, the loop shows a step-like approach to saturation. A plausible interpretation is a two-stage saturation process of an antiferromagnetically coupled bilayer: one moment dominates at low magnetic fields, while at higher fields the antiferromagnetic coupling is overcome and the magnetic moments align, leading to a second increase in $M$ toward full saturation. Using the magnetic free energy of the stack \mbox{Cr(10)/Gd(5)/Co(3)}, we can estimate the interface coupling energy as $J_\mathrm{int} = B_\mathrm{sat}(T) \min(M_\mathrm{Gd}(T) t_\mathrm{Gd}, M_\mathrm{Co}(T) t_\mathrm{Co}) \approx 10\,\mathrm{mJ}/\mathrm{m^2}$. Numerical solutions of the free-energy model confirm this value, see the Appendix.

\subsection{Anomalous Hall Effect Measurements}
\begin{figure}[t]
  \centering
  \includegraphics[width=\linewidth]{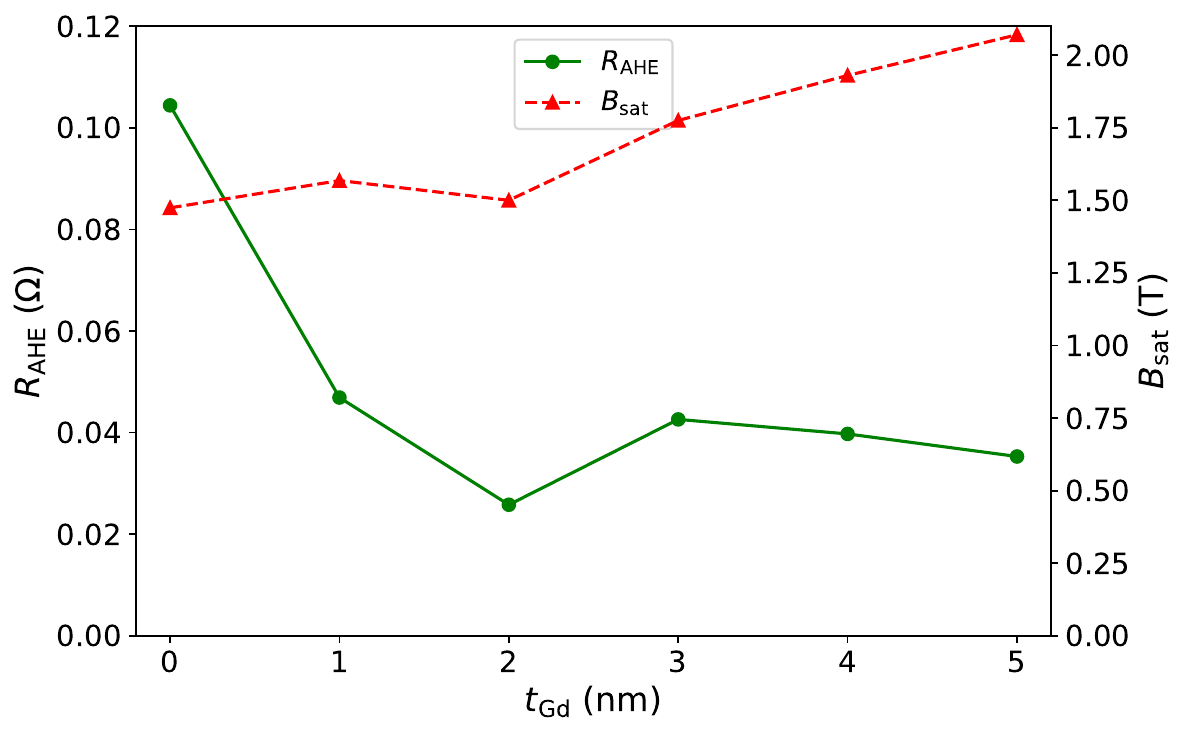}
  \caption{Anomalous Hall effect measurements of the \mbox{Co(3)/Gd($x$)/Cr(10)} stacks with $x = 0\ldots5$~nm, together with the corresponding saturation field $B_\mathrm{sat}$.}
  \label{fig:RAHEBsat}
\end{figure}

With increasing Gd thickness, $R_\mathrm{AHE}$ decreases strongly from \mbox{$0.104~\Omega$} for the Gd-free sample and reaches a minimum at \mbox{$t_\mathrm{Gd}=2~\mathrm{nm}$}, see ~Fig.\ref{fig:RAHEBsat}. In contrast, $B_\mathrm{sat}$ shows an overall increase with increasing Gd thickness. This indicates that the insertion of Gd strongly modifies both the AHE response and the magnetic field scale required for out-of-plane saturation. The increase of the saturation field contradicts the apparent reduction of the magnetization at room temperature, which would lead to a reduced demagnetizing field and therefore a smaller saturation field, $B_\mathrm{sat} = \mu_0 M$ in the absence of additional anisotropy fields. While the saturation field of the sample without a Gd interlayer is in perfect agreement with the measured saturation magnetization, the increase of $B_\mathrm{sat}$ can be explained with the magnetic free energy of the stack. A solution in the strong-coupling limit where the Co moment rotates out-of-plane and keeps the (small) Gd moment locked antiparallel gives
\begin{equation}
B_\mathrm{sat}^\perp = \mu_0 \frac{M_\mathrm{Co}^2 t_\mathrm{Co}+ M_\mathrm{Gd}^2 t_\mathrm{Gd}}{\left| M_\mathrm{Co} t_\mathrm{Co} - M_\mathrm{Gd} t_\mathrm{Gd} \right|}.
\end{equation}
This equation directly predicts the observed increase of perpendicular saturation; a fit to the data of Figure \ref{fig:RAHEBsat} indicates $M_\mathrm{Gd} \approx 160$\,kA/m at room temperature. A similar fit to the data of the inverted Cr / Gd / Co stacks (not shown) indicates  $M_\mathrm{Gd} \approx 320$\,kA/m, i.e. a much stronger intermixing and thereby higher effective Curie temperature of the Gd layer and larger interlayer coupling are inferred. This is consistent with the previously discussed higher compensation temperatures in Cr / Gd / Co stacks. The general case allowing for canting of the layer magnetizations and including the interlayer exchange explicitly requires a numerical solution of the model equations. The numerical solutions also indicates that a strong coupling is essential to explain the increase of the saturation field with increasing Gd thickness. In contrast, with weak coupling, a spin-flop state emerges and gradually pulls the Gd and Co magnetizations parallel to the out-of-plane field. In this case, the saturation field is essentially independent of the Gd magnetization. More details on the model analysis are given in the Appendix.
Additional temperature-dependent measurements (not shown) on selected samples demonstrate a substantial enhancement of the perpendicular saturation field beyond 4\,T, which can only be explained by taking a finite interlayer coupling and canted states into account.

\subsection{Ferromagnetic Resonance Measurements}
\begin{figure}[t]
\centering
\includegraphics[width=\linewidth]{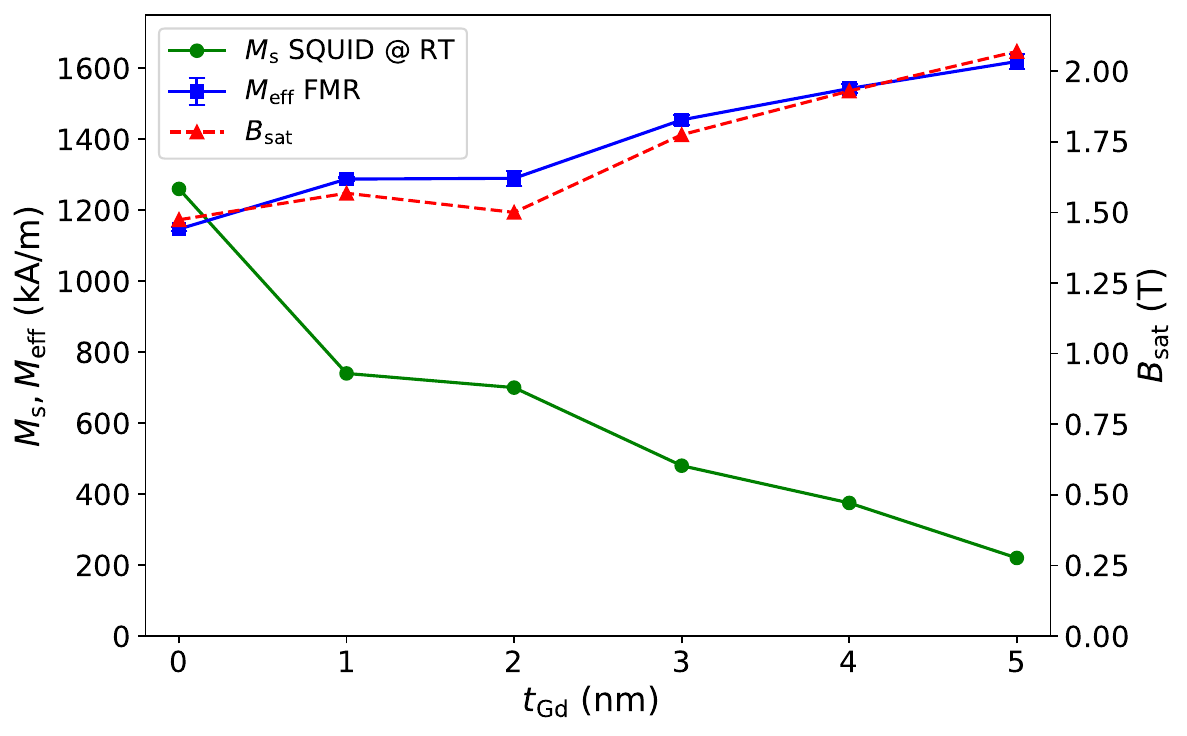}
\caption{Comparison of the saturation magnetization obtained from SQUID magnetometry, the effective magnetization extracted from FMR measurements, and the saturation field for the Co(3)/Gd($x$)/Cr(10) samples. While $M_\mathrm{s}$ decreases with increasing Gd thickness, both $M_\mathrm{eff}$ and $B_\mathrm{sat}$ increase, indicating an enhanced in-plane anisotropy contribution to the FMR response. The y-axis ticks for $M_\mathrm{s}$, $M_\mathrm{eff}$ (kA/m) and $B_\mathrm{sat}$ (T) are aligned according to $B=\mu_0 M$.}
\label{fig:SQUID_FMR_AHE}
\end{figure}

Effective magnetization values of the Co / Gd / Cr multilayers were obtained from FMR measurements and are displayed in Fig.~\ref{fig:SQUID_FMR_AHE}. They increase with increasing Gd thickness. This contrasts with the reduction of the saturation magnetization $M_\mathrm{s}$ determined by SQUID magnetometry, but is in perfect agreement with the increased perpendicular saturation field. As both methods effectively measure the same quantity, the perpendicular saturation field $\mu_0 M_\mathrm{eff} = B_\mathrm{sat}^\perp$, this is a reassuring result for the correctness of our understanding of the out-of-plane saturation field. The decrease in small-field magnetization with increasing Gd thickness indicates that the insertion of Gd substantially modifies the average magnetic properties of the layer stack, which are related to antiparallel Gd--Co coupling and intermixing at the Co/Gd interface.

\subsection{TEM Analysis}
\label{sec:tem}
\begin{figure}[t]
  \centering
  \includegraphics[width=\columnwidth]{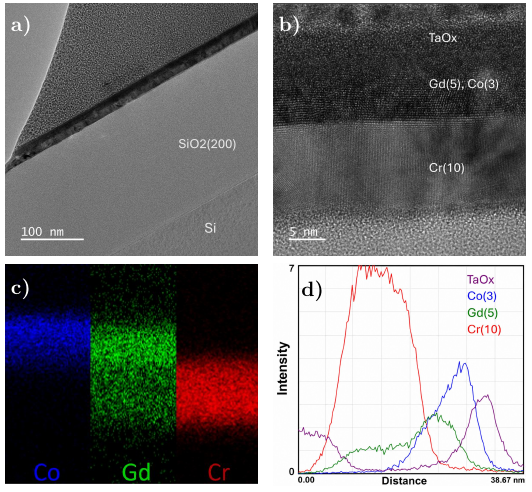}
  \caption{Cross-sectional TEM and EDX analysis of a representative
\mbox{Si/SiO$_2$(200)/Cr(10)/Gd(5)/Co(3)/TaOx} sample. (a) Overview TEM image of the full layer stack on the substrate. (b) Higher-magnification TEM image highlighting the multilayer sequence. (c,d) EDX elemental map and line scan of the \mbox{Cr/Gd/Co} trilayer. The EDX data reveal intermixing of Gd into the FM \mbox{Co} layer. The apparent Gd signal in the Cr layer is likely an EDX artifact, since the Cr K$\beta$ line lies close to the Gd L$\alpha$ line and can be misassigned during peak deconvolution.}
  \label{fig:TEM}
\end{figure}

To understand the magnetic properties of the multilayer stacks, we performed cross-sectional TEM/EDX on the \mbox{Si/SiO$_2$(200)/Cr(10)/Gd(5)/Co(3)/TaOx} sample (Fig.~\ref{fig:TEM}). The overview TEM image in Fig.~\ref{fig:TEM}(a) confirms a continuous thin-film stack on the \mbox{Si/SiO$_2$} substrate, while the high-resolution image in Fig.~\ref{fig:TEM}(b) resolves the \mbox{Cr(10)} layer and the \mbox{Gd(5)/Co(3)} region beneath the TaOx cap. The corresponding EDX elemental map and line scan [Fig.~\ref{fig:TEM}(c,d)] reveal a pronounced intermixing of Gd and Co. The apparent Gd signal within the Cr layer is an artifact of the overlap of the Cr K$\beta$ line and the Gd L$\alpha$ line and can lead to misassignment during peak deconvolution.

This observation of intermixing supports an alloyed, probably amorphous (Co-Gd) magnetic region, consistent with the reduced net magnetization at room temperature. The alloy region has an elevated Curie temperature compared to the Gd bulk Curie temperature. At the same time, the Gd magnetic moments will still be coupled antiparallel to the Co moments, thus leading to a possibly complicated sperimagnetic state. 

In addition, the high-resolution TEM image [Fig.~\ref{fig:TEM}(b)] is consistent with the expected structure across the stack. The thermally grown SiO$_2$ layer appears amorphous, as indicated by the absence of periodic fringes. The Cr(10) layer exhibits locally visible lattice fringes and grain-like contrast, indicative of polycrystalline (nanocrystalline) growth.  At the Cr/Gd interface, lattice fringes with a larger spacing are visible and indicate partial crystallization of the Gd layer, where the bond lengths of ca. 3.6\,\AA{} give rise to the much larger fringe separation compared to Cr with a bond length of 2.57\,\AA{}. In contrast, the upper \mbox{Gd(5)/Co(3)} region shows no clear lattice fringes and more diffuse contrast, consistent with strong intermixing and increased structural disorder with no clear crystallization of the Co film. The TaOx cap appears amorphous, as expected for an oxidized Ta layer.

To summarize the structural and magnetic analysis, our results rule out a ``naive'' layer-by-layer magnetization model based on well-separated Co and Gd layers.

\subsection{Harmonic Hall Analysis}

The above analysis shows that both $R_\mathrm{AHE}$ and $B_\mathrm{sat}$ of Eq. \ref{EQU:V2w_decomp} have a well-defined meaning and the harmonic Hall analysis can be used as a linear-response method to a small electrical perturbation of the FM layer magnetization. However, the analysis also shows that the magnetic layer is in an intermediate state between a homogenous alloy and well-separated layers. Because of the high resistivity of the Gd layer $\rho_{\mathrm{Gd}}\approx215\,\mu \Omega \mathrm{cm}$, there is little current flow in that layer and the electrical response stems largely from the Co layer. 

\begin{figure*}[ht]
  \centering
  \includegraphics[width=\textwidth]{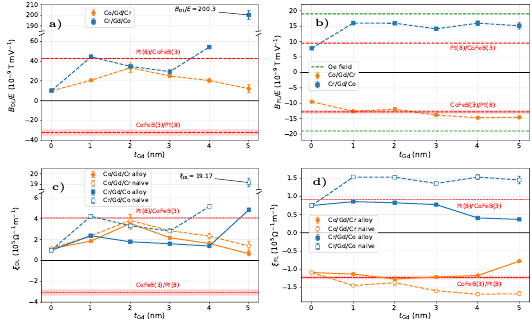}
 \caption{Effective DL (a) and FL (b) fields per applied electric field of the \mbox{Co(3)/Gd($x$)/Cr(10)} and \mbox{Cr(10)/Gd($x$)/Co(3)} series with \mbox{$x = 0 \ldots 5$\,nm}, determined from SHH measurements. Reference samples and Oersted field contribution are included for comparison. (c) DL torque efficiency $\xi_{\mathrm{DL}}^{E}$ with largest magnitude in $\xi_{\mathrm{DL}}^{E}$ for \mbox{Cr(10)/Gd(5)/Co(3)}. Note that a broken y-axis is used to display this outlier without compressing the remaining data. (d) FL torque efficiency $\xi_{\mathrm{FL}}^{E}$.
Both torque efficiencies are normalized to the applied electric field $E$. Results are shown for both the naive and the alloy model.}
  \label{fig:Xi}
\end{figure*}

For an initial analysis of the DL and FL torques, we quantify the torque amplitudes in terms of the effective magnetic fields per applied electric field, $B_{\mathrm{DL, FL}}/E$, see Figs.~\ref{fig:Xi}(a) and (b). Unlike the corresponding torque efficiencies $\xi_{\mathrm{DL/FL}}^{E}$ (Eq. \ref{eq:xiDLE}), these quantities do not require normalization by the FM layer thickness or the saturation magnetization. Depending on the stack order, the measured DL torque is overall of a similar magnitude as in our $\mathrm{Pt}(8)/\mathrm{CoFeB}(3)$ reference sample. An outlier in the more intermixed Cr / Gd / Co series displays a $500\,\%$ larger DL torque compared to the reference. A clear peak around 2\,nm Gd in the Co / Gd / Cr series is visible, which supports the orbital-to-spin conversion scenario of Ref. \cite{2022SalaGiantOrbitalHall}.

The most striking observation in our study, which has not been documented so far in the literature to the best of our knowledge, is the lack of sign reversal of $B_\textrm{DL}/E$ upon reversing the stack order in all cases, including $t_\mathrm{Gd} = 0$\,nm. It can not be explained with a self torque of the stack without Cr(10), since a Co(3)/Gd(2) reference sample exhibited only small self-torque effective fields per applied electrical field of $B_\mathrm{DL}/E=1.262 \times 10^{-9}$\,TmV$^{-1}$ and  $B_\mathrm{FL}/E=0.305 \times 10^{-9}$\,TmV$^{-1}$. The self torque efficiencies are $\xi_{\mathrm{DL}}^{E}=(0.145\pm0.020)\times10^{5}\,\Omega^{-1}\mathrm{m}^{-1}$ and $\xi_{\mathrm{FL}}^{E}=(-0.035\pm0.007)\times10^{5}\,\Omega^{-1}\mathrm{m}^{-1}$.

In contrast, the FL torque changes sign, consistent with a simple Oersted field and an interfacial (Rashba/Edelstein-like) contribution. The former can be estimated as $B_\mathrm{Oe} = \mu_0 j t_\mathrm{Cr} / 2$ and using $j = \sigma E$ we get the Oersted efficiency as $B_\mathrm{Oe} / E = \mu_0 \sigma t_\mathrm{Cr} / 2 \approx 19 \cdot 10^{-9}$\,Tm/V using the measured resistivity of Cr $\rho = 1/\sigma \approx 33\,\mu\Omega$cm. This value is significantly larger than the measured FL effective field, $B_{\mathrm{FL}}/E$, for both stack sequences, as shown in Fig.~\ref{fig:Xi}(b). This indicates an additional FL torque with a sign opposite to that of the Oersted field. 

For the conventional analysis of the DL and FL torque efficiencies, we consider both models for the FM in [Eq.~(\ref{eq:xiDLE})]. On the one hand, we assume the alloy model with
\begin{equation}
t_{\mathrm{FM}} = t_{\mathrm{Co}} + t_{\mathrm{Gd}},
\end{equation}
and $M_s$ from the SQUID data for each $t_{\mathrm{Gd}}$ normalized to the same thickness. For comparison, we apply the \emph{naive model} that uses the measured $M_s$ for \mbox{$t_{\mathrm{Gd}}=0$\,nm} and assumes a fixed FM thickness $t_{\mathrm{FM}}=t_{\mathrm{Co}}$, ignoring the magnetic contribution of the Gd layer.

Fig.~\ref{fig:Xi}(c) and (d) summarizes the torque efficiencies normalized to the electric field $E$ for both stack sequences. In the naive model, we observe $\xi_{\mathrm{DL/FL}}^{E}$ being generally substantially larger compared to the alloy model. In the Cr/Gd/Co sequence, the naive model shows a substantial $\xi_{\mathrm{DL}}^{E}$  increase over the alloy model [Fig.~\ref{fig:Xi}(c)]. In contrast, in the Co/Gd/Cr sequence, a small increase of the naive model over the alloy model is found. This sequence also shows a clear extremum in $\xi_\mathrm{DL}$ at $t_\mathrm{Gd} = 2\,\mathrm{nm}$, which reaches nearly the same value as the Pt/CFB reference sample. Our results on the Co/Gd/Cr sequence are fully consistent with the previous observation by Sala \textit{et al.} (see Fig. 5 of Ref. \cite{2022SalaGiantOrbitalHall}), although we find the extremum at $t_{Gd} = 2\,\mathrm{nm}$, whereas they found it at $t_{Gd} = 3\,\mathrm{nm}$. Also numerically our results are practically identical for Co/Cr and for the peak value similar to the SHC of Pt.

The difference between the two magnetic layer models amounts to more than a factor of two in the Co / Gd / Cr sequence and up to a factor of five in the Cr / Gd / Co sequence, where the alloy model always indicates the smaller values. For a quantitative analysis of the orbital Hall effect in rare-earth / transition-metal systems, it is thus mandatory to perform a careful analysis of the magnetic properties of the detection layer.

\section{Discussion}
The abscence of the usually expected sign-reversal of the DL torque upon stack inversion is clearly the most outstanding result of this study. It was carefully tested against typical spin Hall reference systems, which show the sign reversal as usual with an otherwise identical measurement; our standardized packaging (Fig. \ref{fig:sampleInPackage}) and fully automatic switching between Hall bars ensures the reproducibility of the sample alignment with the magnetic field and current. We have also ruled out a large self-torque of the Co / Gd pair, which would not change sign under stack reversal. Thus, we are left with an unexplained lack of inversion symmetry of the stacks. The injected orbital polarization at the two Cr interfaces flips sign, i.e. $\mathbf{L} \rightarrow \mathbf{-L}$. With an interfacial orbital-to-spin conversion process, the oriented interface normal vector flips sign under stack reversal  $\mathbf{n} \rightarrow \mathbf{-n}$. A crystal-field or hybridization-sensitive orbital-to-spin conversion Hamiltonian could have the correct symmetry, i.e. $H_{LS} = \lambda_\mathrm{eff}(\mathbf{n}) \mathbf{L}\cdot\mathbf{S}$ with $\lambda_\mathrm{eff}(-\mathbf{n}) = -\lambda_\mathrm{eff}(\mathbf{n})$.

\FloatBarrier 
\section{Conclusion}
In summary, we investigated thickness-dependent torque generation in Cr/Gd/Co and Co/Gd/Cr heterostructures by combining SHH measurements with SQUID magnetometry, FMR, and cross-sectional TEM/EDX analysis. SQUID magnetometry reveals a strong reduction of the net magnetization with increasing Gd thickness and the emergence of temperature-dependent magnetic compensation points, indicating ferrimagnetic coupling between Co and Gd. In parallel, TEM/EDX shows pronounced intermixing in the Co-Gd region, demonstrating that the magnetic layer cannot be treated as a simple sequence of well-separated Co and Gd layers.

This intermixing strongly affects the extraction of torque efficiencies. A naive analysis based on a fixed Co thickness substantially overestimates the DL torque efficiency, whereas an alloy model using an effective magnetic thickness $t_\mathrm{FM}=t_\mathrm{Co}+t_\mathrm{Gd}$ and the SQUID-derived magnetization provides a more physically meaningful description. Within this framework, the largest DL torque efficiency is observed for the Cr(10)/Gd(5)/Co(3) stack, exceeding the Pt/CoFeB reference. The FL torque changes sign upon stack inversion, consistent with an interfacial contribution, whereas the DL torque does not reverse sign within the SHH analysis.

These results highlight that orbital-to-spin conversion in Cr/Gd/Co heterostructures is governed not only by the nominal layer sequence but also by interfacial mixing and the resulting ferrimagnetic alloy formation. Correlating electrical torque measurements with magnetic and structural characterization is therefore essential for quantitatively interpreting orbital-torque efficiencies in Co/Gd heterostructures.

\appendix
\section{Coupled-macrospin model}
\label{app:macrospin}

\begin{figure}[t]
  \includegraphics[width=8.6cm]{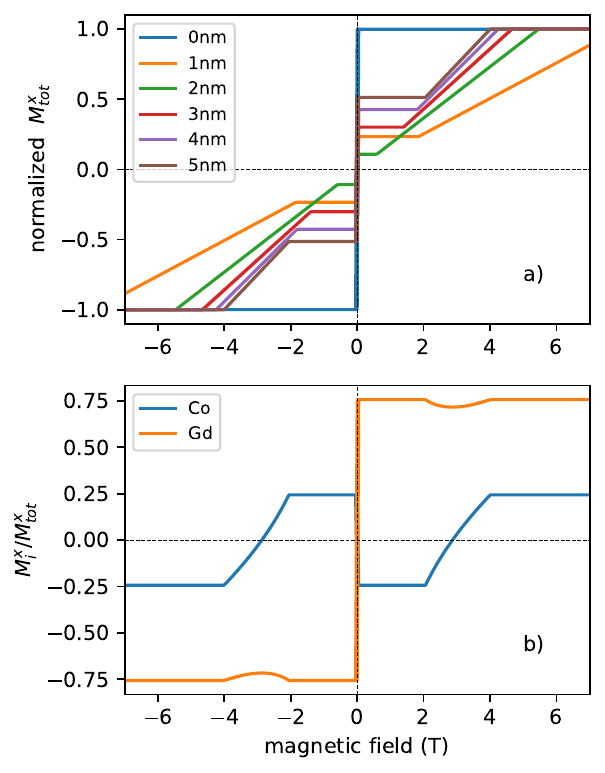}
 \caption{a) Normalized $x$-component of the total magnetization for Gd film thicknesses between 0\,nm and 5\,nm for in-plane magnetic field sweeps along $x$. Parameters are $J_\mathrm{AF} = 10$\,mJ/m$^2$, $M_\mathrm{Co} = 1100$\,kA/m, $M_\mathrm{Gd} = 2050$\,kA/m, $t_\mathrm{Co} = 3$\,nm. The Gd magnetization resembles the value at 5\,K. b) For $t_\mathrm{Gd} = 5$\,nm, the layer-resolved $x$-components of the magnetizations.}
  \label{fig:macrospin_ip_5K}
\end{figure}

\begin{figure}[t]
  \includegraphics[width=8.6cm]{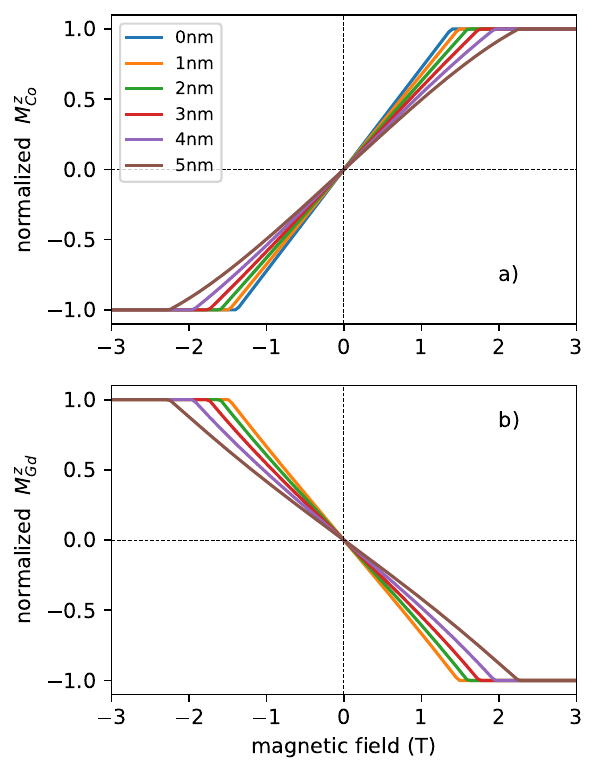}
 \caption{Normalized $z$-component of the Co layer magnetization (a) and Gd layer magnetization (b) for Gd film thicknesses between 0\,nm and 5\,nm for out-of-plane magnetic field sweeps along $z$. Parameters are $J_\mathrm{AF} = 10$\,mJ/m$^2$, $M_\mathrm{Co} = 1100$\,kA/m, $M_\mathrm{Gd} = 180$\,kA/m, $t_\mathrm{Co} = 3$\,nm The Gd magnetization representsthe room temperature (ca. 300\,K) state.}
  \label{fig:macrospin_oop_300K}
\end{figure}

The Co and Gd layers are treated as uniform macrospins with areal moments
\(m_i=M_i t_i\). For an out-of-plane field, the magnetization directions are parametrized as
\begin{equation}
\mathbf{u}_i=(\sin\theta_i,0,\cos\theta_i),
\end{equation}
and the energy per unit area is
\begin{align}
\mathcal{E}_{\perp}
={}&
\frac{1}{2}\sum_i m_i B_{d,i}\cos^2\theta_i
-B\sum_i m_i\cos\theta_i
\nonumber\\
&+
J_{\mathrm{AF}}
\cos(\theta_{\mathrm{Co}}-\theta_{\mathrm{Gd}}),
\label{eq:energy_oop}
\end{align}
where \(B_{d,i}\simeq\mu_0M_i\) is the thin-film demagnetizing field and \(J_{\mathrm{AF}}>0\) favors antiparallel alignment.

For an in-plane field applied along \(x\), the moments rotate azimuthally,
\begin{equation}
\mathbf{u}_i=(\cos\phi_i,\sin\phi_i,0),
\end{equation}
giving
\begin{align}
\mathcal{E}_{\parallel}
={}&
-B\sum_i m_i\cos\phi_i
+
J_{\mathrm{AF}}
\cos(\phi_{\mathrm{Co}}-\phi_{\mathrm{Gd}}).
\label{eq:energy_ip}
\end{align}
The demagnetizing energy is constant in this geometry and has therefore been omitted.

At each applied field, the equilibrium orientations are obtained by solving
\begin{equation}
\frac{\partial\mathcal{E}}{\partial q_{\mathrm{Co}}}
=
\frac{\partial\mathcal{E}}{\partial q_{\mathrm{Gd}}}
=0,
\end{equation}
where \(q_i=\theta_i\) or \(\phi_i\) for the out-of-plane or in-plane geometry, respectively. Only stationary solutions with a positive-semidefinite energy Hessian are retained. Field sweeps are calculated by continuously following the stable solution closest to that obtained at the preceding field value.

We present two main results from the numerical solutions, the normalized total magnetization in-plane loops at 5\,K (Fig. \ref{fig:macrospin_ip_5K}) and the normalized Co out-of-plane component at 300\,K (Fig. \ref{fig:macrospin_oop_300K}). While we compare the former to the SQUID measurements of Fig. \ref{fig:SQUIDCoGdCrMH}, clearly showing the step-like increase of the magnetization when both film magnetizations are pulled into the same direction, the latter demonstrate the increase of the out-of-plane saturation field (compare with Fig. \ref{fig:SQUID_FMR_AHE}) due to the antiparallel coupling of the two layers. Notably, with weak coupling, the saturation field would reduce with increasing Gd film thickness. Thus, the increase of the saturation field is direct evidence for the strong antiparallel coupling which persists at room temperature.

\FloatBarrier 
\section*{AUTHOR DECLARATIONS}
\subsection*{Conflict of Interest}
The authors have no conflict to disclose.
\subsection*{Author Contributions}
ChatGPT-5.6 Sol was used to implement the numerical solution of the coupled macrospin model.\\
\textbf{Tiago de Oliveira Schneider}: Conceptualization (equal); Data curation (lead); Formal analysis (lead); Investigation (lead); Methodology (equal); Project administration (lead); Resources (lead); Software (equal); Supervision (lead); Validation (lead); Visualization (lead); Writing -- original draft (lead); Writing -- reviewing \& editing (equal).\\
\textbf{Michel Heidkamp}: Investigation (supporting).\\
\textbf{Luana Caron}: Investigation (equal).\\
\textbf{Inga Ennen}: Investigation (equal).\\
\textbf{Matthias Opel}: Investigation (equal).\\
\textbf{Alexey Arzumanov}: Investigation (supporting).\\
\textbf{Markus Meinert}: Conceptualization (equal); Data curation (supporting); Formal analysis (supporting); Funding acquisition (lead); Methodology (equal); Project administration (supporting); Software (equal); Supervision (supporting); Validation (supporting); Writing -- reviewing \& editing (equal).
\bibliographystyle{apsrev4-2}
\section*{REFERENCES}
\bibliography{references}

\end{document}